\documentclass{INTERSPEECH2023}
\usepackage[T1]{fontenc}
\usepackage{wrapfig}
\usepackage{subfigure}
\usepackage{graphicx}
\usepackage{bm}
\usepackage{makecell}
\usepackage{epstopdf}

\interspeechcameraready

\title{A Multi-Scale Attentive Transformer for Multi-Instrument Symbolic Music Generation}
\name{Xipin Wei$^{1,2}$, Junhui Chen$^1$, Zirui Zheng$^1$, Li Guo$^{1*}$, Lantian Li$^1$, Dong Wang$^2$\thanks{
This work was supported by the National Natural Science Foundation of China under Grant No.62171250.}}
\address{
  $^1$School of Artificial Intelligence, Beijing University of Posts and Telecommunications, China \\
  $^2$Center for Speech and Language Technologies, BNRist, Tsinghua University, China}
\email{$^*$Corresponding authors:~guoli@bupt.edu.cn}

\begin{document}

\maketitle
\begin{abstract}

Recently, multi-instrument music generation has become a hot topic.
Different from  single-instrument generation, multi-instrument generation needs to consider inter-track harmony besides intra-track coherence.
This is usually achieved by composing note segments from different instruments into a signal sequence.
This composition could be on different scales, such as note, bar, or track.
Most existing work focuses on a particular scale, leading to a shortage in modeling music with diverse temporal and track dependencies.

This paper proposes a multi-scale attentive Transformer model to improve the quality of multi-instrument generation.
We first employ multiple Transformer decoders to learn multi-instrument representations of different scales
and then design an attentive mechanism to fuse the multi-scale information.
Experiments conducted on SOD and LMD datasets show that our model improves both quantitative and qualitative performance
compared to models based on single-scale information.
The source code and some generated samples can be found at https://github.com/HaRry-qaq/MSAT.

\end{abstract}
\noindent\textbf{Index Terms}: music generation, multi-instrument, multi-scale, Transformer

\section{Introduction}

Recently, symbolic music generation with deep learning techniques has received much attention~\cite{lyu2015polyphonic,zhu2018xiaoice,dong2018musegan}.
According to the number of instruments in the generated music, the generation task can be divided
into single-instrument generation (SIG) and multi-instrument generation (MIG)~\cite{briot2020deep}.
Most of the research so far focused on SIG tasks, trying to produce coherent
note sequences. Impressive progress has been achieved in this field, particularly in generating
long music~\cite{oore2017learning,huang2018music,waite2016generating,yang2017midinet,hsiao2021compound,hawthorne2022general}.

In recent years, more research attention has been moved to multi-instrument generation tasks~\cite{zhou2018bandnet,donahue2019lakhnes,ens2020mmm}.
Compared to single-instrument music, multi-instrument music is more expressive by flexibly utilizing diverse instruments
collaboratively. However, MIG is much more complex than SIG as it needs to consider
inter-track harmony besides intra-track coherence.
This indicates that MIG is more rewarding but also more challenging.

Two important issues in MIG research are (1) how to compose events of multiple instruments into \emph{a single sequence},
or multi-instrument event serialization; and (2)
how to design a machine learning model for learning such sequences, making sure that both the intra-track and inter-track dependency
can be well represented. These two issues are clearly correlated and should be considered together.

Zhou et al.~\cite{zhou2018bandnet} proposed BandNet. They encoded the multi-instrument music (Beatles' songs) via
a note-level zig-zag scan strategy, i.e., from left-to-right (time dimension) and
top-to-bottom (track dimension), and then used a 3-layer LSTM-RNN to learn the encoded sequence.
Donahue et al.~\cite{donahue2019lakhnes} designed an event-based representation for multi-instrument generation called LakhNES.
They designed 631 events about half time-related events and half note-related events, and serialized events from
multiple instruments according to the occurrence time. They used a Transformer to model the merged event sequence.
Ens et al.~\cite{ens2020mmm} presented a Multi-Track Music Machine (MMM), which preserves the whole sequence of musical events for each 
instrument and concatenates sequences of multiple instruments 
into a single sequence. They also used a Transformer model, with the hypothesis that the attention mechanism can learn both the inter-track and the intra-track  dependency if the information of instrument and position is appropriately encoded.
Recently, Dong et al.~\cite{dong2022multitrack} introduced a Multitrack Music Transformer (MMT). They defined a new
representation that supports multiple instruments. Specifically, they encapsulated a music event and related information, 
such as beat, position, pitch, duration, and instrument, as a tuple, and serialized the events in time sequence.
The tuple sequence was modeled by Transformer. Since the instrument information has been included in the tuple,
learning the inter-track and intra-track dependencies is supposed to be easier for the Transformer model.

\begin{figure*}[!htp]
	\centering
	\includegraphics[width=0.52\linewidth]{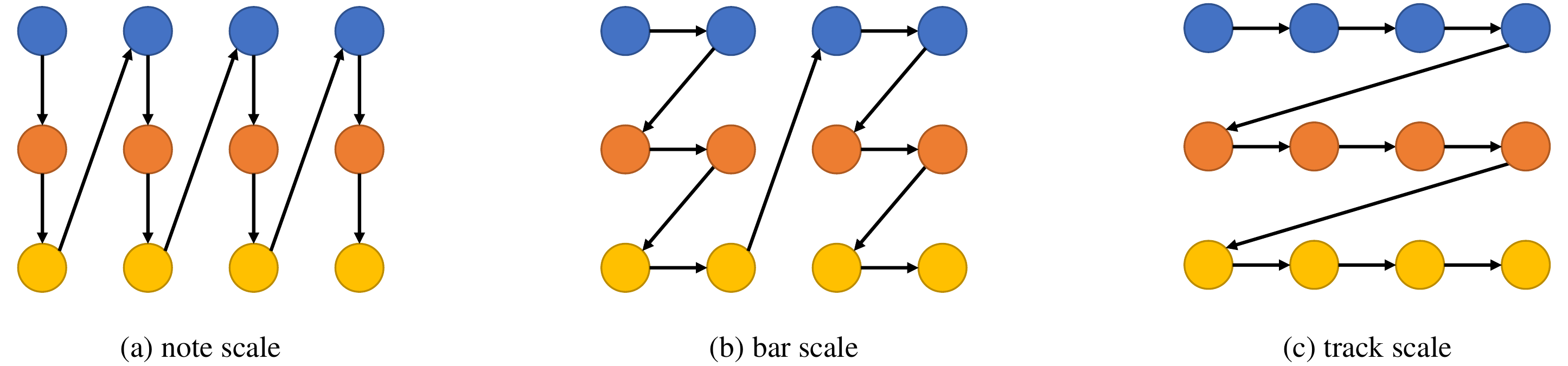}
    \vspace{-2mm}
	\caption{Multi-instrument composition at (a) note scale; (b) bar scale; (c) track scale. Each row represents an instrument and each column represents a time step.}
	\label{fig:scale}
\end{figure*}

The composition can be conducted at different scales, i.e.,
at which position the serialization can move from one instrument to another.
Figure~\ref{fig:scale} illustrates the serialization process at the note-level, the bar-level, and the track-level.
From this perspective, BandNet, LakhNES, and MMT are based on note-level composition,
while MMM is based on track-level composition. Different scales of composition offer
different information scopes for the model in both training and inference, and require different positioning
strategies when using Transformer as the backbone.

Most of the existing studies are based on a single-scale composition, which
could have prevented the model from learning diverse music.
This is because the music of different genres tends to exhibit inter-track and intra-track dependencies on different scales.
For example, \emph{Canon} music is usually composed in strict counterpoint by bar, so bar-level information is sufficient for different instruments to look at each other.
For the \emph{piano concerto}, it requires more attention to the musical information of the piano track,
so it is more appropriate for other instruments to have track-level attention on the piano.
For the \emph{Jazz} music with the theme of improvisation, higher frequency variations in rhythm
and accent~\cite{gioia2011history} are often observed between notes, so note-level cross-instrument attention is most useful.


\begin{figure*}[!htp]
	\centering
	\includegraphics[width=0.96\linewidth]{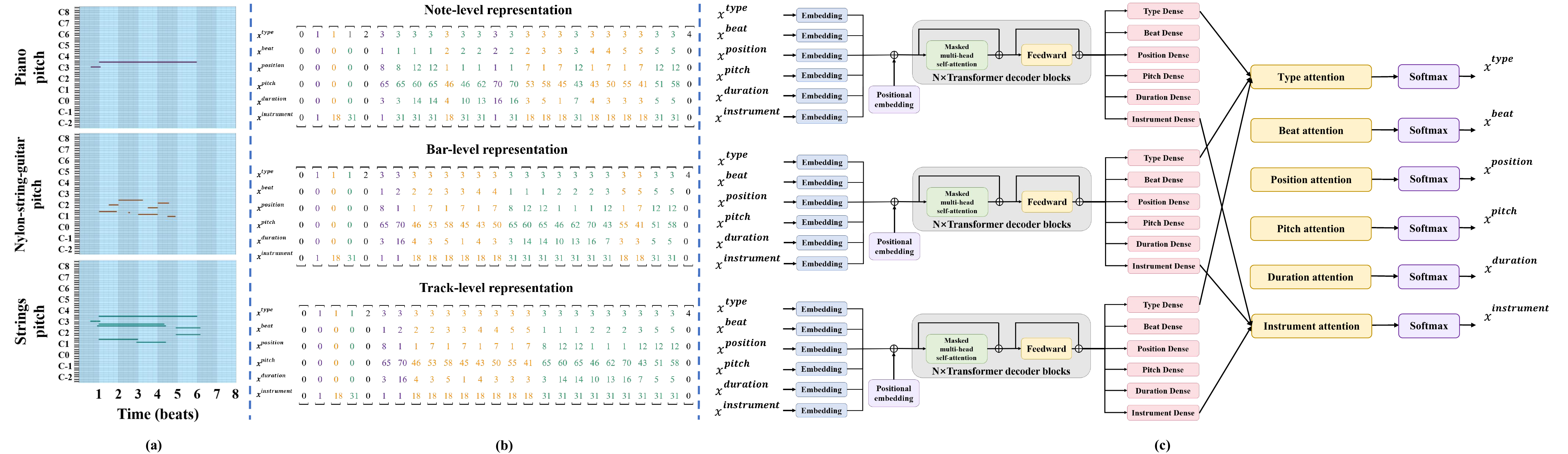}
   \vspace{-2mm}
	\caption{Illustration of our proposed multi-scale representation model - (a) An example of the first eight beats of a song shown as a multi-track piano roll. (b) The song is encoded by three scales of representations, from top to bottom are note-level, bar-level, and track-level respectively. (c) The structure of the proposed multi-scale attentive Transformer model.}
	\label{fig:model}
   \vspace{-2mm}
\end{figure*}


In this paper, we propose a multi-scale attentive Transformer model (MSAT) to
leverage the advantages of composition methods at different scales.
We first employ different Transformer decoders to learn multi-instrument representations of different scales and then design an attentive mechanism to fuse the multi-scale information.
Although it seems Transformer can learn the whole sequence no matter at which scale the
sequence is composed, we found the scale is crucial as the decoder is autoregressive so can only see partial sequence.
Our experiments are conducted on two popular datasets, SOD and LMD.
Experimental results with both the objective evaluation and the subjective listening test
show that our proposed multi-scale learning model achieves consistent performance improvement compared to models based
on single-scale composition.

\section{Multi-instrument MIDI representation}
\label{sec:rep}

In this paper, we encode multi-instrument events following the MMT format~\cite{dong2022multitrack}.
For a music piece, it is represented as a sequence of events $x=(x_1, ..., x_n)$,
where each event $x_i$ is encoded as a tuple of six tokens:
\begin{equation}
\vspace{-1mm}
\label{eq:tuple}
  ( x_{i}^{type}, x_{i}^{beat}, x_{i}^{position}, x_{i}^{pitch}, x_{i}^{duration}, x_{i}^{instrument})
\vspace{-1mm}
\end{equation}
\noindent The \emph{type} token specifies the function of the tuple, e.g., start-of-song, start-of-notes, end-of-song, etc.
The remaining tokens represent the beat of the event, the position in the beat, the note and its duration, and the instrument.
Note that since the instrument information is involved in each event, this format is suitable for composition at flexible scales.
More details about the MMT format can be found in~\cite{dong2022multitrack}.

We then compose the events from multiple instruments into a multi-instrument event sequence by
serializing them at either the note level, bar level, or track level. This results in three multi-instrument
representations, denoted by MMT-note, MMT-bar, and MMT-track. 
Figure~\ref{fig:model}(a) presents an example of a song.
This song consists of three tracks corresponding to three instruments.
The representations at the three levels are shown in Figure~\ref{fig:model}(b).

It should be highlighted that composition at different scales possesses its own
pros and cons. For example, note-level composition
can better describe the correlation of two or more main instruments at a particular time,
while the track-level composition is better at learning the long-term dependency between
a main instrument and a couple of auxiliary instruments.




\section{Multi-scale attentive Transformer}

In this section, we will present our proposed multi-scale attentive Transformer model (MSAT).
This model is based on a Transformer decoder~\cite{liu2018generating,brown2020language}
that inherits from the MMT model~\cite{dong2022multitrack}.
Compared to MMT which only supports the input of single-scale representations,
we design a separate Transformer decoder to learn the representation of each scale, and
introduce token-wise cross-scale attention to fuse information at different scales.

\subsection{Multi-head Transformer decoders}

We duplicate the Transformer used in the MMT model~\cite{dong2022multitrack} three times, each
corresponding to one of the three composition scales: note, bar, and track.
Each decoder consists of two components.
The first one is the pre-processing component which contains a token-wise embedding layer and a positional encoding operation~\cite{vaswani2017attention}.
And the second one is a series of Transformer decoder blocks, each containing a masked multi-head self-attention layer and a feed-forward layer.

\subsection{Token-wise cross-scale attention}

To fuse the multi-scale information, we design two kinds of token-wise cross-scale attention methods.
As shown in Figure~\ref{fig:model}(c),
the Transformer decoder at each scale outputs the embedding vector of the input event.
A linear projection is then used to decompose the event embedding to six token embeddings,
each corresponding to a particular token shown in Eq.(\ref{eq:tuple}).
Then for each token, an attentive fusion is designed to merge the embeddings of that token
at the three scales.
We experimented with two designs for cross-scale attention, which differ in how to compute the attention weights.


\subsubsection{Global attention}

In the global attention approach, we train a set of attention weights $\omega_n, \omega_b, \omega_t$
to scale the token embeddings from the note-level, bar-level, and track-level representations, and
the weights are used for all the events. Note each token type has its own set of weights.

Take the instrument token as an example.
We define the token embeddings as $h_n^{inst}, h_b^{inst}, h_t^{inst}$,
which are derived from note-level, bar-level, and track-level representations respectively.
The attention coefficients are then computed by Softmax:
\vspace{-1mm}
$$ \alpha_{i} = \frac{\text{exp}(\omega_{i})}{\sum_{k \in \{n,b,t\}}\text{exp}(\omega_{k})}, $$
\noindent where $\omega_n, \omega_b, \omega_t$ are learnable.\noindent The fused instrument embedding is computed as:
\vspace{-1mm}
$$ h^{inst} =\sum_{i \in \{n,b,t\}} \alpha_i ~ h^{inst}_{i}. $$

\subsubsection{Local attention}

Unlike global attention, local attention focuses on computing the attention weights according to the token embeddings.
Also, take the instrument token as an example.
We define a learnable instrument-dependent attention matrix $ W \in \mathbb{R}^{3 \times N} $,
where $N$ is the dimensionality of the instrument embeddings.
First compute the attention score $ \hat\alpha_{\{n,b,t\}} $ as follows:
\vspace{-1mm}
$$ \hat\alpha_{\{n,b,t\}} = W \cdot  [~h_{n}^{inst}, h_{b}^{inst}, h_{t}^{inst}~], $$
\noindent where $ \hat{\alpha}_{\{n, b, t\}} \in \mathbb{R}^{3 \times 1} $.
Then the fused instrument embedding is computed by:
\vspace{-1mm}
$$ h^{inst} = \sum_{i \in \{n,b,t\}} \alpha_{i} ~ h_i^{inst}, $$
\noindent where
\vspace{-1mm}
$$ \alpha_{i} = \frac{\text{exp}(\hat{\alpha}_{i})}{\sum_{k \in \{n,b,t\}}\text{exp}(\hat{\alpha}_{k})}. $$
\noindent Note that the attention weights are different at each time, which is why it is named `local' attention.

\section{Experiments}

\subsection{Data}

Two publicly available multi-instrument datasets were used in our experiments,
Symbolic Orchestral Database (SOD)~\cite{crestel2018database} and Lakh MIDI Dataset (LMD)~ \cite{raffel2016learning}.
We first discarded those instruments without pitch, such as drums, and selected songs with 4/4 time.
We used MusPy~\cite{dong2020muspy} to process the data. Finally, we obtained 5,742 songs from SOD and 5,916 songs from LMD.
For both datasets, we split 80\% of the data for training, 10\% for validation, and 10\% for testing.

\subsection{Model Settings}

Firstly, for the single-scale approach, we directly employed the MMT model~\cite{dong2022multitrack}
and trained three single-scale models, denoted by MMT-note, MMT-bar, and MMT-track.

For the multi-scale approach, although the input involves three representations of different
scales, there should be only one single-scale output (target).
We chose the bar-level representation as the target and found better performance than choosing others.
Moreover, to make the model training more stable, the decoders of MMT-note and MMT-track are employed in the multi-scale model (weight fixed) as the decoders for the note-level and track-level representations.
Therefore, only the bar-level decoder and the attentive module need to be trained.
We refer to the model trained with the global attentive fusion as MSAT-GA and the local attentive fusion as MSAT-LA.

In our experiments, all the training configurations are the same as MMT,
and the models are trained to minimize the sum of the cross-entropy losses of different tokens in an autoregressive way.


To evaluate the generation models, we designed two generation tasks as follows:

\begin{itemize}
  \item \emph{Task 1. Instrument-informed generation}: a sequence of instrument codes are
  extracted from a ground true music and given to the model.
  The model then generates the note sequence.
  Music samples generated in this task are used for the objective evaluation.
  \item \emph{Task 2. N-beats continuation}: all instrument and note events in the first $N$ beats are provided to the model.
  The model then generates subsequent note events that continue the input music.
  In our experiments, $N$ is set to 16. The generated music samples are used for the subjective listening test.
\end{itemize}

It is worth mentioning that we did not adopt the unconditioned generation in our experiment. We are more concerned with the quality rather than the flexibility of different models in a multi-instrument generation. Therefore, during the generation, we will pre-define the desired combination of 
instruments (Task 1) or the `sample' for generation (Task 2).

\begin{table*}[t]
  \centering
  \caption{Performance comparison of our proposed model against three single-scale models.}
  \vspace{-2mm}
	\label{tab:result}
	\scalebox{0.75}{
	\begin{tabular}{llcccccccccc}
		\toprule
		&& \makecell{pitch \\ class entropy} & \makecell{scale \\ consistency} & \makecell{groove \\ consistency} & \makecell{inter-instrument \\ similarity}  &  \makecell{instrument \\ consistency} \\
		\midrule
		\textbf{SOD}           & Ground truth  &  2.813          &  92.81           &  93.99           &  0.215            &  1.000   \\
		\midrule
		single-scale
                            &  MMT-note     &  2.017          &  96.16           &  \textbf{95.88}   &  0.793           &  0.593   \\
                            &  MMT-bar      &  2.195          &  96.87           &   97.97           &  0.647           &  0.720   \\
                            &  MMT-track    &  1.936          &  96.32           &   98.55           &  0.719           &  0.380   \\
     \midrule
     multi-scale            &  MSAT-LA      &  2.164          &  96.61           &   98.08           &  \textbf{0.611}  &  \textbf{0.747}   \\
                            &  MSAT-GA      &  \textbf{2.403} &  \textbf{92.11}  &   97.36           &  0.614           &  0.695   \\
		\midrule
		\midrule
		\textbf{LMD}           &  Ground truth  &  2.423         &  96.01           &  94.94            &  0.398        &  1.000   \\
     \midrule
     single-scale           &   MMT-note     &  1.590         &   98.30          &  \textbf{97.43}   &  0.827           &  0.493    \\
                            &   MMT-bar      &  1.625         &   99.00          &  97.49            &  0.621           &  0.604    \\
                            &   MMT-track    &  \textbf{1.900}  &  97.55        &  98.68             &  0.634           &  0.281    \\
		\midrule
     multi-scale            &  MSAT-LA       &  1.796         &  98.28           &  97.51            &  \textbf{0.578}  &  \textbf{0.668}    \\
                            &  MSAT-GA       &  1.873         &\textbf{96.98}    &  98.11            &  0.642           &  0.539    \\
		\bottomrule
	\end{tabular}}
\vspace{-2mm}
\end{table*}

\subsection{Objective evaluation}

\subsubsection{Metrics}

Several metrics are designed to evaluate the performance of the models on the instrument-informed generation task.
To evaluate intra-track coherence,
we follow~\cite{mogren2016c,wu2020jazz} and measure the pitch class entropy,
scale consistency, groove consistency, and empty measure rate for each instrument.

For inter-track harmony, we compute the inter-instrument similarity
which is the standard deviation of the pitch class entropy of different instruments.
According to music theory, the higher the similarity between the pitch distributions of two
instruments, the more instrumental harmonious the music will sound.

Finally, we define instrument consistency as the correlation between the desired
number of instruments and the number of instruments in the generated music in terms of bar.
This metric reflects the controllability of the model on instruments.

For these metrics, we hope they are close to the values of the ground truth music where
the instrument codes are extracted.

\subsubsection{Results}

For each model, 399 and 544 music pieces were generated using ground-truth samples from SOD and LMD respectively.
Table~\ref{tab:result} shows the evaluation results.
Firstly, let us compare the three single-scale models.
Combining different types of evaluation metrics, we can observe MMT-bar outperforms MMT-note and MMT-track as a whole.
This observation suggests that
the bar-level representation takes into account both the temporal coherence of each instrument and the harmonic
correlation between different instruments, thus suitable for multi-instrument music generation.

Secondly, our proposed MSAT model outperforms the single-scale competitor MMT-bar in almost all the evaluation metrics,
with either local attention or global attention. The two MSAT models obtain the best performance in most of the metrics.
All the results demonstrated that the multi-scale learning approach allows the model to discover more inter-track and intra-track dependency
from representations composed at different scales.

\subsection{Subjective listening test}

To further assess the quality of the music samples generated by our proposed model,
we invited 28 participants to conduct a listening test. Most of the participants are music producers or music practitioners.
Each participant is required to listen to 10 groups of tests, and
each group contains three music samples that were generated by MMT-bar, MSAT-GA, and MSAT-LA in the 16-beat continuation task.
The participants were asked to choose their favorite generation, considering the following criteria:
\begin{itemize}
\item \emph{Coherence}: Is it temporally coherent? Is the rhythm steady? Are there many out-of-context notes?
\item \emph{Richness}: Is it rich and diverse in musical textures? Are there any improper repetitions and variations? Is it too boring?
\item \emph{Arrangement}: Are the instruments used reasonably? Are the instruments arranged properly?
\item \emph{Quality}: Overall quality of the sample.
\end{itemize}

Results are presented in Table~\ref{tab:human}.
Firstly, we found that MSAT-GA and MSAT-LA gained more votes on Richness and Arrangement
while losing some votes on Coherence.
This indicates that MSAT-GA/LA obtained better inter-track harmony while sacrificing a bit of intra-track coherence.
Besides, it can be observed that from the overall preference, MSAT-LA received the most votes,
suggesting that this model is favored in terms of overall performance.

\begin{table}[!htp]
 \centering
 \caption{Subjective listening test results by preference voting.}
 \vspace{-2mm}
 \label{tab:human}
 \scalebox{0.75}{
 \begin{tabular}{l|cccc|c}
   \toprule
     Votes                &  Coherence  &  Richness  &  Arrangement  &  Quality  &  Overall      \\
   \midrule
        MMT-bar           &   \textbf{102} &  84        &    93        &    93      &    372        \\
        MSAT-GA           &    85       &   \textbf{99} &    88        &    86      &    358        \\
        MSAT-LA           &    93       &   97        &  \textbf{99}   &   \textbf{101}   &    \textbf{390}    \\
  \bottomrule
  \end{tabular}}
\vspace{-2mm}
\end{table}

\subsection{Ablation study}

\begin{figure}[!htp]
	\centering
   \vspace{-1.5mm}
	\includegraphics[width=0.71\linewidth]{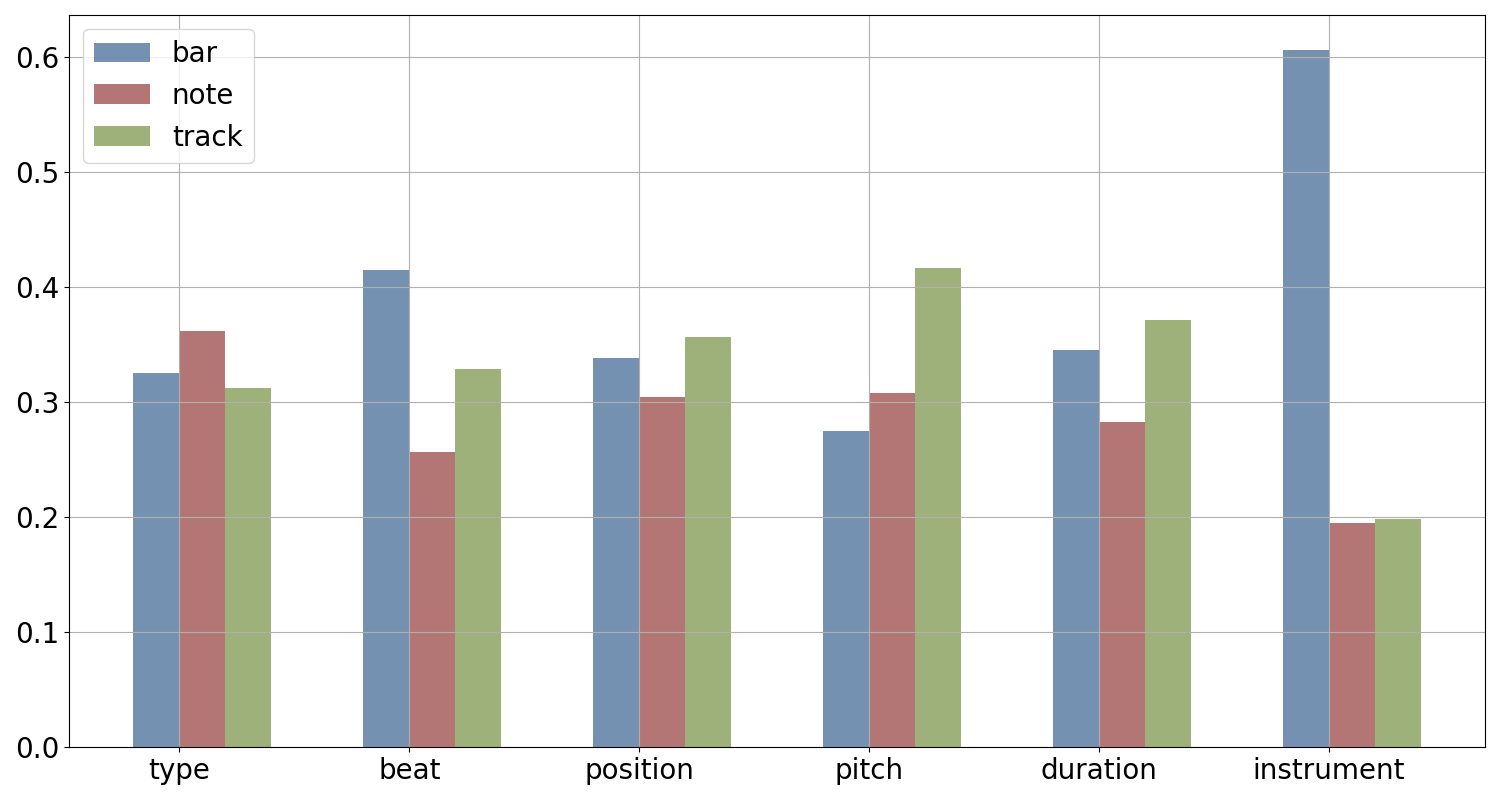}
   \vspace{-2mm}
	\caption{Global attention weights.}
	\label{fig:att}
   \vspace{-2mm}
\end{figure}

In this section, we try to gain a deeper understanding of how the cross-scale attention
looks into the representations composed at different scales.
We plot the attention weights $\omega_n, \omega_b, \omega_t$ in the
global attention model, as shown in Figure~\ref{fig:att}. Note these weights are
token-wised.

Firstly, for the \emph{type} token, the attention weights are almost the same among the three scales.
This is reasonable since this token is not very informative to the composition scale.
Secondly, for \emph{beat} and \emph{instrument}, the bar-level scale
dominates the attention weights. We conjecture that these two tokens represent the main difference between the composition schemes, and because the bar-level representation is the target in the model learning, it would be easy for the model to converge if it got more information from the bar-level input.
Finally, for \emph{position}, \emph{pitch}, and \emph{duration}, the information of the
track-level composition seems more prominent, indicating that the long-term information is also
important to guide inter-instrument harmony.

The above observations validate our multi-scale learning approach: it is not a single-scale input that dominates in the
learning; instead, information from the inputs of the three scales seems all important.

\section{Conclusions}

This paper proposed a multi-scale attentive Transformer model (MSAT) to improve the quality of multi-instrument music generation.
Instead of learning from the multi-instrument representations composed at a single scale, MSAT
learns representations composed at three scales, i.e., note, bar, or track.
Experimental results on the objective evaluation and subjective listening test showed that our proposed MSAT model
obtained better performance than the models trained with representations composed at any single scale.
In the future, we will consider a more flexible structure to achieve multi-scale learning and
improve intra-track coherence in multi-instrument music generation.

\newpage

\bibliographystyle{IEEEtran}
\bibliography{mybib}

\end{document}